# When Do Traditional and Causal Decomposition Methods Diverge? A Practical Guide for Studying Health Disparities

Soojin Park, Chioun Lee, and Su Yeon Kim

University of California, Riverside

Author Note

The authors acknowledge the assistance of Gemini for grammar checking and editing.
Correspondence concerning this article should be addressed to Soojin Park, School of Education, University of California-Riverside, CA 92521, USA. Email: soojinp@ucr.edu

Abstract

Identifying malleable factors that can reduce social disparities is a central objective among researchers across disciplines. Traditionally, researchers have relied on the difference-in-coefficients and Kitagawa-Oaxaca-Blinder frameworks. More recently, methods grounded in the potential outcomes framework have emerged. While these methods share the same goal of identifying drivers of disparity, they frequently yield divergent results depending on the underlying confounding structures and research settings. Despite these significant differences, applied researchers lack clear guidance on selecting appropriate methods for their specific research contexts. To address this gap, this study provides a systematic review and offer an intuitive guidance through Directed Acyclic Graphs and comparative simulation studies. We begin by reviewing each method assuming no unmeasured confounding, which is often violated in observational settings. Consequently, we extend our analysis to two realistic scenarios: 1) unmeasured confounding exists in the relationship between intermediate confounders and the mediator, and 2) unmeasured confounding exists in the relationship between the mediator and the outcome. Finally, we illustrate these recommendations through a case study examining the role of educational attainment in explaining racial disparities in later-life cognition.

*Keywords:* Difference-in-coefficients; Kitakawa-Oaxaca-Blinder; Causal Decomposition; Unmeasured Confounding; Health Disparity; Social Disparity.

**When Do Traditional and Causal Decomposition Methods Diverge?**
**A Practical Guide for Studying Health Disparities**

## 1. Introduction

Persistent disparities in health and social outcomes across race/ethnicity, sex/gender, and class remain a central concern in the US. For example, there are enormous racial disparities in later-life cognitive function (e.g. Weuve et al., 2018), with recent estimates of cognitive life expectancy indicating that the number of years living with dementia is more than twice as high for Blacks than Whites (Garcia et al., 2019). Since one cannot modify ascribed characteristics like race/ethnicity and sex, the interest of researchers and policymakers often centers on identifying malleable risk factors that may play a role in reducing such disparities. For example, a substantial share of the Black-White differential in cognitive function (e.g., memory, executive reasoning, and speed of processing) in the US may result from disparities in socioeconomic status (SES), including educational attainment, occupation, income, and wealth (e.g. Sheftel, Goldman, Pebley, Pratt, & Park, 2025; Lee, Glei, & Park, 2024). These risk factors are referred to as 'mediators' since they are believed to lie between social groups (race) and an outcome (cognition).

To identify mediators that can help reduce such disparities, researchers have traditionally relied on linear modeling frameworks, such as the difference-in-coefficients (DIC) approach (Olkin & Finn, 1995) and the Kitagawa-Oaxaca-Blinder decomposition (KOB) (Kitagawa, 1955; Oaxaca, 1973; Blinder, 1973). More recently, there has been a significant shift toward potential outcome-based methods (Imai, Keele, & Tingley, 2010; VanderWeele & Hernán, 2012; VanderWeele & Robinson, 2014; Jackson & VanderWeele, 2018). While these methods share the common goal of identifying mediators that may reduce social disparities, they often yield divergent results even when applied to the same data due to fundamental differences in underlying confounding structures and modeling assumptions. Despite these contradictions, applied researchers have limited guidance on selecting appropriate methods for their specific causal context. This gap is critical as choosing an inappropriate method can produce spurious findings, potentially leading to implementation of ineffective interventions or policies.

As the main goals of this study, therefore, we first offer a clear explanation for conditions under which the results across methods may diverge. We then provide evidence-based guidance for choosing robust methods tailored to specific research contexts. While existing literature has contrasted different potential outcomes approaches (Nguyen, Schmid, & Stuart, 2021), introduced new causal methodologies by highlighting their advantages over traditional mediation (Jackson & VanderWeele, 2018), or focused on comparing specific estimators within a single potential outcome method (Park, Kang, & Lee, 2024), few studies have systematically compared these emerging causal methods with the traditional linear models. Most recently, Opacic, Wei, and Zhou (2025) provided a systematic review comparing the KOB and potential outcome-based approaches. However, their comparison has relied heavily on algebraic derivations, which may limit their accessibility to applied researchers. A targeted review of *Prevention Science* articles published from 2016 to 2026 found that among application studies estimating mediation effects, 65 of 70 (92.9%) relied on traditional regression-based approaches, whereas only five (7.1%) used counterfactual causal approaches (see Supplementary Material A). This under-utilization of causal approaches highlights the need for a practical guide comparing these methodologies. This paper aims to bridge this gap by comparing them across realistic and varied scenarios, using graphical representations and simulation studies.

The remainder of this article is organized as follows. In Section 2, we introduce a running example to develop a Directed Acyclic Graph (DAG, Pearl, 2009) representing the presumed data-generating process. Section 3 is then provides a review of the four decomposition methods. Within this review, we demonstrate how each framework defines causal estimands (i.e., the effects of interest) depending on different conditions. We supplement this theoretical review with Monte Carlo simulation studies to illustrate the conditions under which these methods diverge. This section concludes with a practical guidance for choosing optimal methods based on two fundamental diagnostic questions.

We then extend this analysis in Section 4 by investigating more realistic scenarios involving unmeasured confounding in various relationships. This evaluation focuses specifically on observational settings where the mediator is not randomly assigned. Given that unmeasured confounding is almost ubiquitous challenge in such contexts, we evaluate how the bias impact the validity of each method. In Section 5, we demonstrate the practical implications of these divergent results by applying the different methods to the case study. Finally, we conclude with a discussion of our findings and recommendations for applied researchers in Section 6.

## 2. Motivating Example

To motivate the study, we ask the question: to what extent does educational attainment explain racial disparities in cognition later in life?" Education, in particular, plays a key role in explaining racial disparities in cognition in later life (Walsemann, Fisk, Farina, Abbruzzi, & Ailshire, 2023). That is possibly because education directly affects brain structure early in life, promoting reserve capacity (Cabeza et al., 2018), which enables individuals to better withstand age-related neural decline and can delay the onset of cognitive impairment (Mungas et al., 2018). Thus, interventions that promote equity in education between Blacks and Whites may yield a substantial reduction of disparities in cognitive function between the groups.

**Directed Acyclic Graphs.** To address this question, we first clarify the relationships among variables and identify confounders with DAGs. In a DAG, a single-headed arrow $A \to B$ indicates a causal effect of A on B. A bi-directed arrow $A \leftrightarrow B$ indicates an association between A and B, often due to unmeasured confounders $U$. Specifically, $U \to A$ and $U \to B$, and thereby inducing an association between A and B. A box around A indicates conditioning on $A$. For an introduction to drawing DAGs and their graphical rules, see Quimpo and Steiner (2026).

As illustrated in Figure 1A-C, we hypothesize that race $R$ has its effect on cognition $Y$ by two primary pathways: P1) indirect pathways through education $M$, representing the portion of the total disparity explained by racial differences in educational attainment (e.g., $R \to M \to Y$), and P2) direct pathways not mediated by education $M$, representing the residual disparity that persists after accounting for these differences (e.g., $R \to Y$). In the decomposition literature, P1 is referred to as the "explained component" (Oaxaca, 1973) or "disparity reduction" (Jackson & VanderWeele, 2018), while P2 is characterized as the "unexplained component" or "disparity remaining". Notably, while these definitions are conceptually consistent, the structural pathways corresponding to these estimands differ across methods, which we review in detail in Section 3.

The assignment of race ($R$) is non-random. Thus, its relationships with educational attainment ($M$) as well as with cognition ($Y$) are confounded by baseline covariates ($C$). We operationalize $C$ as a vector of predisposing demographic characteristics, such as sex, age, or birth cohort. In the context of our example, we identify age as a critical baseline covariate. In longitudinal aging research, such as the *Health and Retirement Study* (HRS), respondents enter the study at a wide range of ages (i.e., age 51 or older). Given that older adults from racial minority groups are less likely to participate in cognitive assessments (expressed as $C \leftrightarrow R$), and age is a key marker of cognitive decline (expressed as $C \to Y$), controlling age is essential in examining the relationship between race and cognition. In our DAG, $C$ and $R$ are connected by a bi-directed edge rather than a directed arrow, as age does not causally influence on race. Rather, this edge represents a non-causal association that could induce a spurious correlation if left unaddressed.

Furthermore, because educational attainment ($M$) is not randomly assigned, its relationship with cognition ($Y$) is confounded by two sources of confounding: 1) baseline covariates and 2) intermediate confounders. We define intermediate confounders ($X$) as a vector of variables influenced by race ($R$) that subsequently affects both the mediator and the outcome. In our example, we specify three intermediate confounders as childhood socioeconomic status (SES) $X_1$, family instability $X_2$, and childhood health $X_3$, representing early-life adversities $X = (X_1, X_2, X_3)$. These factors are affected by or correlated with race through structural historical processes, including racism or the legacy of slavery (Kaufman, 2008; Jackson & VanderWeele, 2018). Furthermore, these factors serve as strong predictors of their educational attainment and cognitive function.

While it is possible to further decompose the indirect and direct pathways to isolate effects operating through these historical processes (see e.g., Jackson & VanderWeele, 2018; Park, Qin, & Lee, 2024), we remain a parsimonious representation illustrated in Figure 1. This allows us to clearly demonstrate on how the estimands identified by each method manifest differently within a simplified DAG framework. Although parsimonious, this DAG remains theoretically applicable to a broad range of empirical scenarios. Specifically, while baseline covariates $C$ and intermediate confounders $X$ are depicted as individual nodes, they represent vectors of variables. Therefore, researchers can incorporate the appropriate sets of confounders needed by their specific choice of mediator and outcome.

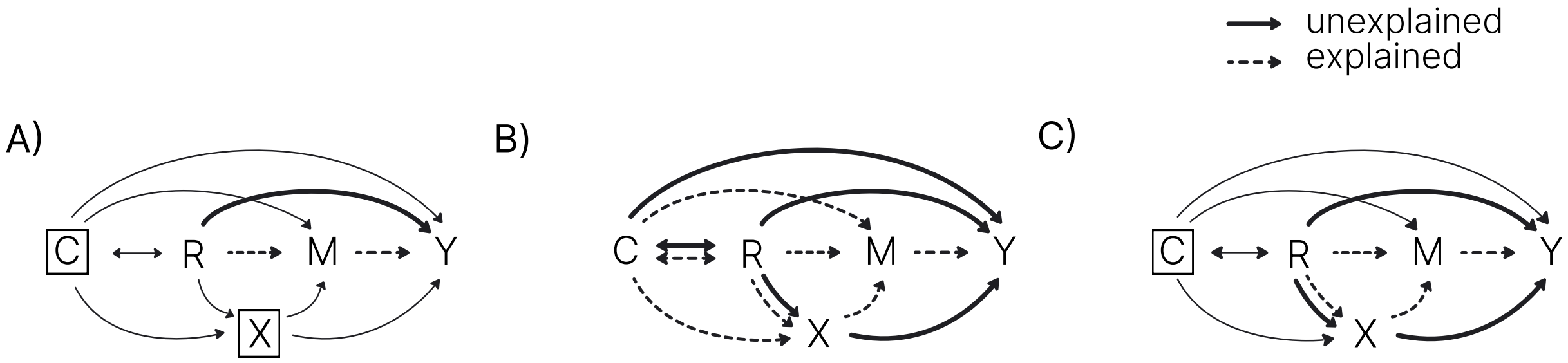

*Figure 1*. DAGs Assuming No Unmeasured Confounders for Each Method

*Note: 1) Dotted arrows indicate the portion of the disparity attributable to differences in the risk factor M. 2) Bold arrows indicate the portion of the disparity not explained by differences in the risk factor M. 3) Standard arrows indicate the pathways block by confounders C or X. 4) A double-headed arrow indicates an association whose causal direction is unknown. 5) A box around a node indicates conditioning on that variable.*

**Data and Measures.** This study utilizes data from the HRS, a nationally representative longitudinal survey of U.S. adults aged 51 and older, conducted biennially since 1992. New cohorts of respondents were added roughly every six years to maintain national representativeness. Core interviews with age-eligible respondents and their spouses have been conducted consistently every two years (Sonnega & Weir, 2014). For this study, we compiled data from four HRS resources: the RAND HRS Longitudinal File, the Childhood Family and Health Aggregated file, the 2020 Tracker File, and the 2017 and 2019 waves of Life History Mail Survey. As our analysis focuses on cognitive function in midlife and later life, we restricted our analytic sample (n = 12,949) to respondents born in 1965 or earlier (late Boomers or older) who completed the LHMS (for details about sample selection, see Lee, Glei, Park, & Weinstein, 2026).

For the key measures, age was measured based on birth year. Race/ethnicity status was measured using self-identified race/ethnicity, categorized as Non-Hispanic White and Non-Hispanic Black. HRS has assessed cognitive function using the Telephone Interview for Cognitive Status (TICS), which includes four tasks: immediate and delayed recall of ten words (each scored from 0 to 10), serial 7s subtraction (scored from 0 to 5), and counting backward (scored from 0 to 2). Scores from these tasks were added together to create a total cognitive score ranging from 0 to 27. As early-life adversities, childhood SES was measured as the average of six standardized indicators (Cronbach's $\alpha = 0.67$). These indicators included a subjective assessment of family financial situation before age 16; whether the family moved due to financial difficulties before age 16; parental education (years of education for both father and mother); and the father's occupation when the respondent was 16 years of age. We also included whether the respondent experienced family instability, measured by parental divorce before age 16, as well as self-rated childhood health before age 15 (coded from 1 = excellent to 5 = poor).

## 3. Review of Methods

### 3.1. Difference-in-Coefficients Method

One traditionally employed approach across disciplines is the DIC approach. It evaluates the explained portion of the disparity by comparing the regression coefficient of race $R$ before and after adjusting for educational attainment $M$. Specifically, we fit the following models:

$$\begin{aligned} E[Y|R,X,C] =& \alpha_0 + \alpha_1 R + \alpha_2 X + \alpha_3 C, \text{ and} \\ E[Y|R,X,C,M] =& \beta_0 + \beta_1 R + \beta_2 X + \beta_3 C + \beta_4 M. \end{aligned} \tag{1}$$

The total disparity is estimated as $\hat{\alpha}_1$, whereas the remaining portion after controlling for education $M$ is $\hat{\beta}_1$. The proportion of the disparity explained by education $M$ is therefore $\hat{\alpha}_1 - \hat{\beta}_1$. This estimator is easy to implement and straightforward to interpret. As shown in Figure 1A, the total disparity $\alpha_1$ is depicted by the combined pathways between bold and dotted arrows. After we adjust for $M$, the remaining disparity $\beta_1$ is shown by the bold arrows; and the disparity explained by $M$, $\alpha_1 - \beta_1$, corresponds to the dotted arrows. The standard errors for these estimates can be obtained by bootstrapping or the delta method (MacKinnon, 2012).

However, the primary advantage of the DIC approach, simplicity, is compromised when the effect of mediator $M$ on the outcome $Y$ varies across social groups $R$. For example, the theory of minority diminished returns (Assari, 2018) suggests that the protective benefits of education (e.g., on cognitive outcomes) may be smaller for Black adults than for White adults. To account for such heterogeneity, we include the interaction term $\beta_5 R \times M$ into the model. In this specification, the remaining disparity is expressed as $\hat{\beta}_1 + \hat{\beta}_5 M$, indicating that the remaining gap is conditional on the level of $M$. Furthermore, the standard DIC estimator is no longer valid when the outcome is categorical (MacKinnon, 2012). In such cases, Karlson–Holm–Breen (KHB; Karlson, Holm, & Breen, 2012) decomposition is a widely adopted approach. It ensures the comparability of coefficients across nested models by accounting for the rescaling of the error variance in nonlinear models. Another common approach is to compare predicted probabilities rather than coefficients across nested nonlinear models, which avoids the latent-scale rescaling problem and facilitates more interpretable comparisons across model specifications (Williams & Jorgensen, 2023).

Perhaps a more crucial issue of this approach is that DIC controls for intermediate confounders ($X$) *when* calculating the total disparity between racial groups. By statistically controlling for variables like early life adversities (e.g., childhood SES) and age, DIC removes them from the analysis of disparity. That is, it assumes the total disparity in a circumstance where age and childhood SES don't differ by race. In Figure 1A, the total disparity excludes any pathways through early life adversities $\boxed{X}$ and age $\boxed{C}$, as shown as the standard arrows. While adjusting for age-related differences ($C$) in the outcome may be appropriate, adjusting for disparities attributable to early life adversities ($X$) is controversial, as those adversities themselves can be manifestations of structural inequality and arguably should be considered part of disparity (Jackson & VanderWeele, 2018).

Therefore, the standard version of the DIC approach is inherently biased in observational disparity studies where intermediate confounders ($X$) almost surely exist. The method tends to underestimates the total racial disparity generated by structural factors by controlling for intermediate confounders in the baseline model. However, the bias in the DIC method would be eliminated if the mediator, educational attainment $M$, were randomly assigned. However, this condition is not feasible within observational studies.

### 3.2. Kitakawa-Oaxaca-Blinder Decomposition

Another widely used linear model approach, particularly in economics and epidemiology, is the KOB decomposition. The standard KOB decomposition decomposes the average difference in the outcome between two groups into two parts: 1) the portion explained by group differences in the average value of predictors ($C, X$, and $M$), and 2) the unexplained portion (or alternatively, the portion explained by differential returns of the predictors). Formally, let $\bar{Z}_r$ denote the average value of variable $Z$ for racial group $R = r$, where $r = \{0, 1\}$. Now, we fit separate outcome models for each group:

$$\begin{aligned}\bar{Y}_1 =& \beta_{01} + \beta_{x1}\bar{X}_1 + \beta_{c1}\bar{C}_1 + \beta_{M_1}\bar{M}_1, \text{ and} \\ \bar{Y}_0 =& \beta_{00} + \beta_{x0}\bar{X}_0 + \beta_{c0}\bar{C}_0 + \beta_{M_0}\bar{M}_0,\end{aligned} \tag{2}$$

where $\beta_{.r}$ represents the vector of regression coefficients estimated within racial group $R = r$. Then the average difference in the outcome between groups $(\bar{Y}_1 - \bar{Y}_0)$ can be expressed as:

$$\begin{aligned}\bar{Y}_1 - \bar{Y}_0 =& \underbrace{\beta_{x1}(\bar{X}_1 - \bar{X}_0) + \beta_{c1}(\bar{C}_1 - \bar{C}_0) + \beta_{M_1}(\bar{M}_1 - \bar{M}_0)}_{\text{Explained by } X,\ C, \text{ and } M} + \\ & \underbrace{(\beta_{01} - \beta_{00}) + (\beta_{x1} - \beta_{x0})\bar{X}_0 + (\beta_{c1} - \beta_{c0})\bar{C}_0 + (\beta_{M_1} - \beta_{M_0})\bar{M}_0}_{\text{Unexplained}}\end{aligned} \tag{3}$$

The first line of equation (3) represents the portion explained by the group differences in the average values of $X$, $C$, and $M$; the second line is the unexplained portion (or the portion explained by differential returns of the predictors). The KOB estimator accommodates differential effects across groups by fitting separate regression models. While it was originally developed for continuous mediators and outcomes, the framework has been extended to address categorical mediators and outcomes through the use of nonlinear link functions (e.g., Yun, 2004; Fairlie, 2005).

Although the KOB framework was originally developed for descriptive purposes, it is instructive to examine the causal interpretations of its constituent components. The portion explained by $M$ (i.e., $\beta_{M_1}(\bar{M}_1 - \bar{M}_0)$) can be interpreted as a valid causal estimand given that the correct set of confounders is controlled for ($X$ and $C$). However, other explained portion, such as the portion attributed to $X$ (or $C$), do not possess a straightforward causal interpretation. This is because the model fitted in equation (3) does not adjust for confounders of the $X \rightarrow Y$ relationship. Specifically, conditioning on $M$ blocks indirect pathways from $X$ to the outcome $Y$ ($X \rightarrow \boxed{M} \rightarrow Y$), leading to overcontrol bias (Pearl, 2009). For example, if the effect of childhood adversities ($X$) on cognition ($Y$) flows largely through education ($M$), controlling for $M$ results in an estimate that understates the importance of childhood adversities $X$. Therefore, researchers should avoid interpreting explained components of $X$ or $C$ as valid causal contributions. Duncan et al. (1968)'s decomposition effectively addresses this issue by sequentially incorporating predictors (e.g., first adjusting for $C$ alone, and followed by both $C$ and $X$). The resulting change in the explained portion is then interpreted as the unique contribution of the newly added predictor. Recently, Opacic et al. (2025) provided a generalized estimation strategy for this sequential approach.

In our study, for comparative purposes, we restrict our focus to components with a well-defined causal interpretation. We define the average difference in cognition between racial groups ($\bar{Y}_1 - \bar{Y}_0$) as the total disparity. The portion of this disparity explained by educational attainment ($M$) is calculated as $\beta_{M_1}(\bar{M}_1 - \bar{M}_0)$. Consequently, The unexplained portion is derived by subtracting the explained disparity from the total disparity.

One limitation of the standard KOB method is that its failure to adjust the total disparity for baseline covariates ($C$) in its current form. Because the method begins with the average difference between groups ($\bar{Y}_1 - \bar{Y}_0$), total disparity incorporates every pathway from race $R$ to cognition $Y$, including causal pathways and backdoor pathways through baseline covariates $C$ (see Figure 1B). This implies that racial differences in cognition due to differences in age $C$ across racial groups is not adjusted for when measuring the total disparity. However, an age-unadjusted Black-White disparity in cognitive function may be misleading, especially when the age distributions differ significantly between racial groups—for example, the sample includes a higher proportion of young Black individuals relative to Whites. Given that cognition tends to decline with age, failing to adjust for age differences across racial groups may underestimate the true extent of racial disparities in cognition at baseline.

The bias of not adjusting for baseline covariates are also folded into both the explained and unexplained portions (represented as dotted and bold arrows, respectively). In the context of our example, an age-unadjusted racial difference in educational attainment ($R \rightarrow M$) would be underestimated (as would the resulting explained portion), given that the Black participants includes a higher proportion of younger individuals compared to the White participants. Likewise, the racial difference in cognition that remain not explained by education is also underestimated due to these differing age distributions between racial groups.

### 3.3. Potential Outcome-Based Methods

The first two methods (DIC and KOB) rely on linear models and the effects of interest cannot be defined without those specific models (Imai et al., 2010). In contrast, potential outcome-based methods provide a model-agnostic approach by defining effects based on the difference between potential outcomes. In addition, potential outcome-based methods focus on the causal interpretation of effects of interest.

**Difference between Natural and Interventional Indirect Effects.** Under this framework, a potential outcome is defined as the outcome that would have been observed for an individual under a specific condition (in our case, a specific levels of educational attainment) (Rubin, 1974). This allows for the formal definition of different scenarios such as: 1) what would the average racial disparity in cognition be if the distribution of educational attainment for Black individuals were shifted to match that of White individuals? and 2) what would the average disparity in cognition be if each Black individual received the level of education they would have attained had they been White instead of Black?

While these two questions may appear similar, they represent a fundamental conceptual difference in causal analysis. The first question involves an intervention at the *population level*: it asks how the disparity would change if we intervene to equalize the educational level between the two groups. This corresponds to a stochastic intervention where Black population is assigned an educational profile that matches that of the White population. In contrast, the second question implies intervention at the *individual level*, which asks what each Black individual's education would have been had their race been different (VanderWeele, Vansteelandt, & Robins, 2014). Unlike the first scenario, it is difficult to link to a real-world intervention, as it requires a change of an immutable characteristic (race) rather than malleable factors such as educational attainment at the individual level.

The first scenario corresponds to analysis based on interventional analogues of natural indirect/direct effects (interventional effects, VanderWeele & Robinson, 2014; Jackson & VanderWeele, 2018) and the second scenario corresponds to analysis based on natural indirect/direct effects (natural effects, Imai et al., 2010; VanderWeele & Hernán, 2012). In light of these distinctions, natural indirect effects are generally better suited for explanatory purposes, such as identifying the underlying mechanisms of a phenomenon or a treatment effect. Conversely, interventional indirect effects are more suitable for policy-oriented interventions as it provides an expected change in an outcome when a mediator is modified in a real-world setting (Nguyen et al., 2021).

Furthermore, this identification of natural indirect effects requires the restrictive assumption of 'no intermediate confounding', which is almost surely violated in observational data. This assumption may be met in experimental designs where the mediator is randomly assigned. However, given the observational nature of our example and the focus of this study (social disparities), we instead focus on interventional indirect effects that offer straightforward definition in the context of disparity research and circumvent the restrictive assumption. Consistent with the existing literature (e.g., Jackson & VanderWeele, 2018; Park, Kang, & Lee, 2024), we refer to the framework that applies interventional in/direct effects to social disparities research as causal decomposition analysis (CDA).

**Definitions.** First, the initial disparity ($\tau$) represents the population-level difference in expected outcomes between racial groups after accounting for baseline covariates. Formally, the initial disparity is defined as:

$$\tau = \sum_c \left( \underbrace{E[Y|R=1, C=c]}_{\text{Conditional mean for Blacks}} - \underbrace{E[Y|R=0, C=c]}_{\text{Conditional mean for Whites}} \right) \underbrace{P(C=c)}_{\text{Marginalizing over C}} \tag{4}$$

In our example, the initial disparity ($\tau$) is defined as the average cognition difference between Blacks and Whites within the same age groups. This initial disparity is then marginalized over the distribution of age to provide population-level estimate that is independent of age-based differences. It is important to note that CDA does not attempt to estimate the causal effect of a social group category, such as race; rather, it focuses on decomposing the observed disparity. Instead of assigning a causal interpretation of immutable factors, the method focuses on estimating the causal effect of specific mediators, such as educational attainment, which are more amenable to intervention (VanderWeele & Robinson, 2014; Lundberg, 2024).

Now imagine a hypothetical intervention that increases Blacks' educational attainment to the same level observed for Whites within the same age group. Disparity reduction ($\delta(1)$) represents the expected change in the outcome for Blacks ($R = 1$) if their mediator distribution were shifted to match that of Whites ($R = 0$) within each level of $C$. Disparity remaining ($\zeta(0)$) represents the remaining average difference in the outcome that would persist even if the mediator distributions were equalized between groups. Formally,

$$\delta(1) = \sum_c \left( \underbrace{E[Y|R=1, C=c]}_{\text{Observed Mean (Blacks)}} - \underbrace{E[Y(G_{m|c}(0))|R=1, C=c]}_{\text{Counterfactual Mean (Blacks)}} \right) \underbrace{P(C=c)}_{\text{Marginalizing over C}} \tag{5}$$

$$\zeta(0) = \sum_c \left( \underbrace{E[Y(G_{m|c}(0))|R=1, C=c]}_{\text{Counterfactual Mean (Blacks)}} - \underbrace{E[Y|R=0, C=c]}_{\text{Observed Mean (Whites)}} \right) \underbrace{P(C=c)}_{\text{Marginalizing over C}} \tag{6}$$

where $G_{m|c}(0)$ denotes a random draw from the mediator distribution of Whites in age group $C = c$. Then, $E[Y(G_{m|c}(0))|R = 1, C = c]$ represents the average potential outcome for Blacks when they had the same mediator distribution of Whites ($R = 0$) within the same age group. Under the linearity, disparity reduction corresponds to the portion of the disparity explained by group differences in the mediator $M$; and disparity remaining corresponds to the unexplained portion.

As shown in Figure 1C, the initial disparity excludes any pathways through age $C$ while still including those through early-life adversities $X$. By adjusting for $C$ while preserving the effects of $X$, the initial disparity specifically captures structural inequalities arising from differential exposure to early-life adversities while controlling for age. Likewise, the unexplained and explained disparities are defined conditional on $C$ while captures structural inequalities due to $X$.

Here, we only introduce a specific intervention: equalizing the distribution of the mediator between groups within the same levels of covariate $C$. However, depending on research questions, alternative intervention strategies can be used, such as 'benchmark' approach, which requires all subjects to adhere to a pre-specified reference level (e.g., requiring high school completion for all). For a detailed discussion on alternative target interventions, we refer readers to Jackson (2020) and Lundberg (2024).

**Identification and Estimation.** However, we cannot observe the potential outcome under the hypothetical intervention. To give a causal interpretation to disparity reduction and disparity remaining estimates, we need to invoke the following assumptions:

A1. **Conditional independence**: This implies that no unmeasured confounding in the mediator-outcome relationship given the group status ($R$), intermediate confounders ($X$), and baseline covariates ($C$).

A2. **Positivity**: $0 < P(R = r \mid C = c)$ and $0 < P(M = m \mid R = 1, X = x, C = c)$. The conditional probability of membership in each racial group is positive across all levels of baseline covariates. In addition, the conditional probability of each observed mediator values for a comparison group (Black) is positive given measured intermediate confounders and baseline covariates.

A3. **Consistency**: The observed outcome should be the same as the potential outcome when the mediator is set to the same level.

In the context of our example, the conditional independence assumption (A1) remains vulnerable even after controlling for the observed variables ($R$, $X$, and $C$). Because educational attainment is not randomly assigned, unmeasured confounding may exist. Hence, we later introduce sensitivity analysis to evaluate the robustness of our estimates against violations of this assumption. The positivity assumption (A2) implies sufficient covariate overlap across racial groups and across mediator values. This is violated if, for instance, there is a lack of representation of the comparison group within specific strata, such as the oldest age groups or in the highest levels of educational attainment. When it is violated, we can trim observations that fall outside the regions of covariate overlap (Imbens & Rubin, 2015). However, a trade-off of this approach is that the results apply to the subpopulation within the trimmed support rather than target population. Lastly, the consistency assumption (A3) requires the observed cognition for an individual with a specific level of education matches to the potential outcome they would have achieved had their education been set to that level by intervention. This assumption would be met if 1) the intervention does not have multiple versions and 2) one person's educational attainment doesn't affect another person's cognitive outcome.

Given the assumptions, the potential outcome is nonparametrically identified, meaning that it can be estimated without any modeling restrictions. Therefore, any model-agnostic estimator can be used to estimate disparity reduction and disparity remaining. By utilizing flexible estimators, such as the imputation estimator (Sudharsanan & Bijlsma, 2021; Park, Qin, & Lee, 2024), the weighting estimator (Jackson, 2020), or doubly-robust estimator (Lundberg, 2024), the CDA approach can effectively handle both categorical and continuous mediators and outcomes and accommodate complex nonlinear effects (such as high-dimensional interactions effects). See Park, Kang, and Lee (2024) for an overview of different CDA estimators.

### 3.4. Modified Linear Methods

If a researcher has a clearly defined causal estimand, can it be estimated using DIC or KOB approach? In this section, we show how DIC and KOB can be modified to estimate the causal estimands defined in equations (5) and (6).

**Modified DIC Method.** Jackson and VanderWeele (2018) introduced a method to estimate the causal estimand through a difference in coefficients approach, even in the presence of intermediate confounders. In addition to the baseline and adjusted models shown in equations (1), this modified DIC method requires fitting the following model: $E[Y|R,C] = \gamma_0 + \gamma_1 R + \gamma_2 C$.

Given the fitted models, the portion of the disparity explained by the difference in $M$ is estimated as $(\hat{\alpha}_1 - \hat{\beta}_1) + (1 - \hat{\beta}_2/\hat{\alpha}_2)(\hat{\gamma}_1 - \hat{\alpha}_1)$. Here, the first component, $(\hat{\alpha}_1 - \hat{\beta}_1)$, represents the difference-in-coefficients estimator from the standard DIC method. The second term serves as an adjustment term that captures the portion of the disparity explained by the intermediate confounder $X$. Specifically, this portion $(\gamma_1 - \alpha_1)$ is scaled by the portion of mediator $(1 - \beta_2/\alpha_2)$. The unexplained portion is derived by subtracting the explained disparity from the total disparity.

While this modified method accurately estimates the previously defined causal estimands under linearity, a potential limitation is that this modification does not extend readily to settings with categorical or discrete outcomes, and to functional forms that deviate from the linear models specified here.

**Modified KOB Decomposition.** Compared to the modified DIC, modifying the KOB decomposition is more straightforward. To adjust the outcome disparity for baseline covariates (e.g., age-adjusted cognitive disparity), covariate-adjusted outcome values can be used instead of raw outcome values. To further adjust for the mediator disparity for baseline covariates (e.g., age-adjusted educational attainment), covariate-adjustment mediator values can be used instead of raw mediator values. Once the covariate-adjusted mediator and outcome are obtained, these values are directly applied to the standard KOB method.

Specifically, the adjusted outcome (or mediator) values can be obtained as follows. First, center the baseline covariates $C$ to their sample means, denoted as $\tilde{C}$. Second, regress the outcome (or mediator) on the group status indicator and baseline covariates $(Y \sim R + \tilde{C})$. Finally, calculate the predicted outcome (or mediator) values. These generated predictions represent the covariate adjusted values.

### 3.5. Simulation Study

Using simulated data, we evaluate the performance of the three aforementioned methods (DIC, KOB, and CDA) in estimating the explained and unexplained portions of a social disparity in a given outcome. We did not include the modified versions of DIC and KOB from these simulations, as they are designed to recover the same true values as CDA under identical model specifications. The simulated data was generated according to the following scenarios: 1) no confounders (both baseline and intermediate), 2) baseline covariates $C$ present, 3) intermediate confounders $X$ present, and 4) both intermediate confounders $X$ and baseline covariates $C$ present.

The first and second scenarios correspond to settings where the mediator is experimentally randomly assigned, effectively eliminating mediator-outcome confounding. In the first scenario, there is no demographic imbalance between racial groups across demographics such as age. Conversely, the second scenario introduces baseline confounding, where the racial groups differ in their demographic distributions, creating an association between race and age.

The third and fourth scenarios correspond to observational settings where the mediator is not randomly assigned, leaving the mediator-outcome relationship susceptible to confounding. Specifically, the third scenario assumes that baseline covariates are balanced across racial groups. Conversely, the fourth scenario introduces baseline confounding, which is frequently encountered in population-based aging studies.

**Data generation.** We generated synthetic data consisting of racial group $R$, mediator $M$, baseline covariate $C$, intermediate confounder $X$, unmeasured confounder $U$, and outcome variable $Y$. The specific data generation procedure is detailed in Supplementary Material B. While our data-generating process utilizes single variables for $C$ and $X$, this simplification does not diminish the generalizability of our findings. The inclusion of multiple variables for $C$ and $X$ does not change the relative performance of the estimators, assuming the estimation models are correctly specified.

We then drew a random sample of $n = 2000$ from the population and applied the estimators reviewed in Sections 3.1-3.3. This procedure was repeated for 500 iterations. To evaluate performance, we compared the averaged estimates against the true values. True values are generated from the causal estimands defined in equations (5) and (6). We also report the 5th and 95th percentiles of the estimates to characterize their empirical variance. For an intuitive understanding, we have bolded the average estimates with the bias is less than 3% of the true value. In the literature (e.g., Flora & Curran, 2004), a threshold $< 5\%$ is widely used to define acceptable relative bias. However, given the sample size of 2000 and the relatively small number of variables in the model, we consider a more conservative threshold of $< 3\%$.

**Simulation Results.** Table 1 shows simulation results that corroborate the pathways shown in Figure 1. In the absence of intermediate confounders $X$ and baseline covariates $C$ (Scenario 1), all three methods yield estimates with negligible bias ($< 1\%$ of the true value) for all three estimands. This demonstrates that when the mediator is randomized and baseline covariates are balanced between groups, all decomposition methods ensure unbiased estimates. However, the KOB method produces biased estimates when baseline covariates ($C$) are strongly associated with group membership (Scenario 2),

Table 1

*Disparity Estimates Assuming No Unmeasured Confounders*

| | DIC | KOB | CDA | True |
|---|---|---|---|---|
| *Without $C$ and $X$* | | | | |
| Total Disparity | **−2.240** [−2.330, −2.160] | **−2.250** [−2.340, −2.150] | **−2.250** [−2.350, −2.150] | −2.249 |
| Unexplained (Remaining) | **−2.000** [−2.080, −1.920] | **−2.000** [−2.090, −1.910] | **−2.000** [−2.100, −1.910] | −2.001 |
| Explained (Reduction) | **−0.244** [−0.286, −0.196] | **−0.247** [−0.288, −0.205] | **−0.248** [−0.306, −0.202] | −0.248 |
| *With $C$ only* | | | | |
| Total Disparity | **−2.240** [−2.370, −2.110] | −2.470 [−2.610, −2.340] | **−2.240** [−2.380, −2.110] | −2.251 |
| Unexplained (Remaining) | **−1.990** [−2.100, −1.870] | −2.440 [−2.590, −2.310] | **−1.990** [−2.130, −1.850] | −1.998 |
| Explained (Reduction) | **−0.255** [−0.327, −0.187] | −0.031 [−0.098, 0.029] | **−0.249** [−0.334, −0.153] | −0.252 |
| *With $X$ only* | | | | |
| Total Disparity | −2.240 [−2.350, −2.150] | **−2.390** [−2.480, −2.270] | **−2.390** [−2.480, −2.280] | −2.387 |
| Unexplained (Remaining) | −2.000 [−2.100, −1.910] | **−2.120** [−2.220, −2.020] | **−2.120** [−2.210, −2.030] | −2.122 |
| Explained (Reduction) | −0.247 [−0.293, −0.201] | **−0.264** [−0.316, −0.221] | **−0.263** [−0.312, −0.216] | −0.265 |
| *With $C$ and $X$* | | | | |
| Total Disparity | −2.260 [−2.360, −2.160] | −2.530 [−2.400, −2.650] | **−2.390** [−2.550, −2.230] | −2.388 |
| Unexplained (Remaining) | −2.010 [−2.100, −1.910] | −2.330 [−2.460, −2.200] | **−2.130** [−2.250, −1.990] | −2.123 |
| Explained (Reduction) | −0.254 [−0.308, −0.205] | −0.200 [−0.259, −0.150] | **−0.261** [−0.373, −0.144] | −0.265 |

*Note:* $C$ refers to the baseline covariate; $X$ refers to the intermediate confounder; DIC = Difference-in-Coefficients; KOB = Kitagawa–Oaxaca–Blinder (Blacks' coefficients as the reference); CDA = Causal Decomposition Analysis. Empirical 95% confidence intervals are shown in brackets. Estimates with small relative bias (less than 3% away from the true value) are bolded.

even when the mediator itself is randomly assigned.

In the presence of intermediate confounders $X$ (Scenario 3), the DIC estimates yield biased estimates. This implies that in observational settings where the mediator is not randomly assigned, the DIC is not an appropriate choice. When both $X$ and $C$ are present (Scenario 4), only the CDA produces estimates with negligible bias ($< 1\%$ of the true value). These results demonstrate that in realistic observational settings, where the mediator is not randomly assigned and baseline covariates are imbalanced between groups, CDA is the only viable methodology for recovering the true disparity estimands.

**Guidelines for Method Selection.** Our simulation results suggest that the choice of decomposition method should be governed by their research setting and confounding structure. Specifically, applied researchers must ask: 1) whether the mediator is randomly assigned? and 2) whether baseline covariates are significantly associated with the group indicator (addressing covariate imbalance). Our specific recommendation is shown in Figure 2. While the recommendations presented in Figure 2 apply to the reviewed methods in their standard form, it is worth noting that a modified KOB decomposition can easily be substituted for CDA.

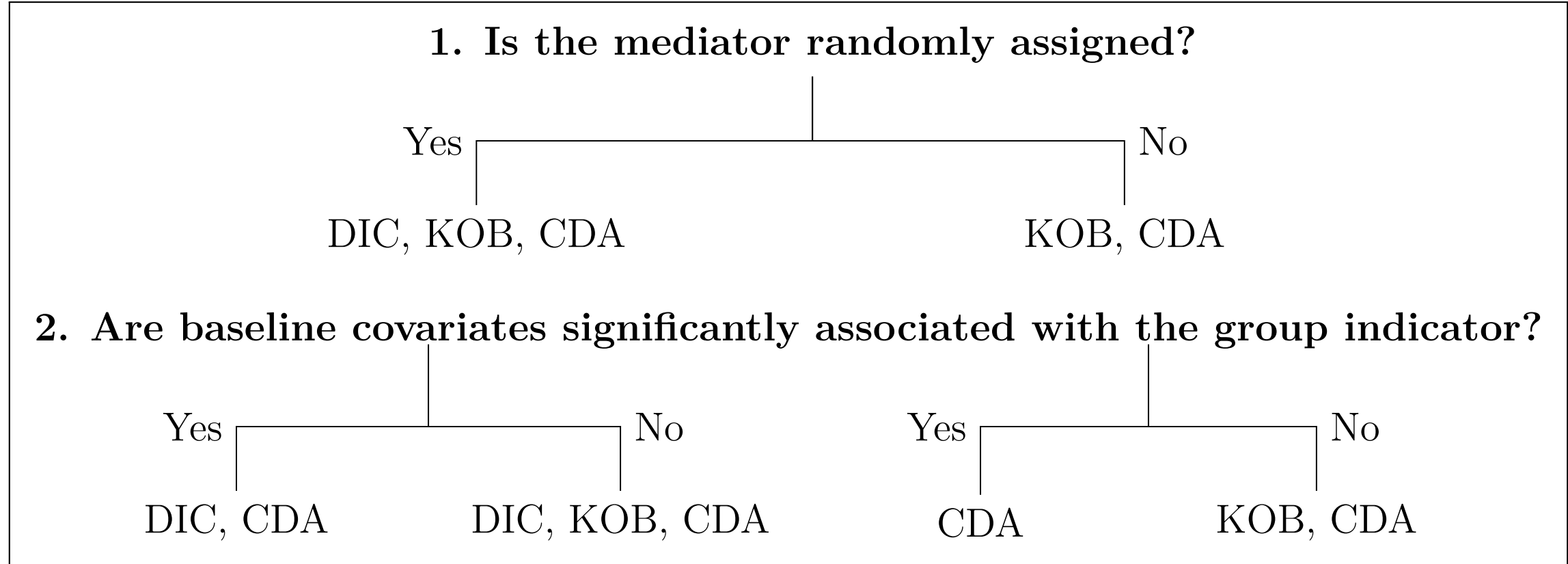

*Figure 2*. Decision Tree Diagram for Method Recommendations

Note: DIC = Difference-in-coefficients; KOB = Kitagawa–Oaxaca–Blinder decomposition; CDA = causal decomposition analysis.

It is noteworthy that this recommendation assumes the estimation models are correctly specified. In practice, with a larger number of baseline covariates and intermediate confounders, the risk of model misspecification increases. While addressing misspecification is beyond the scope of this study, we direct readers to developments of debiased/double machine learning (Chernozhukov et al., 2018) as a way to mitigate these issues and its applications to investigating social disparities (e.g., Lundberg, 2024; Park, Kim, Zheng, & Lee, 2025).

## 4. Considering More Realistic Scenarios

In this section, we focus on an observational setting characterized by potential unmeasured confounding. In a realistic scenario, unmeasured confounding $U$, such as neighborhood environment in early life, may simultaneously influence childhood adversities ($X$), the educational attainment ($M$) and cognition ($Y$). For analytical clarity, we decompose this problem into two distinct scenarios involving an unmeasured confounder $U$ that affects: 1) the mediator-intermediate confounder ($M - X$) relationship and 2) the mediator-outcome ($M - Y$) relationship.

A primary challenge in this assessment is that bias arising from these unmeasured confounding is compounded with the different effects generated across methods. Furthermore, one might contend that the DIC and KOB framework only appear biased within the simulation study (in Section 3.4) because the true effects are generated based on the CDA framework. A researcher may intentionally choose to adjust for factors like childhood SES $X_1$ (as is done in the DIC approach) to isolate a pure effect of race that is independent of SES effects. To accommodate these research objectives and precisely isolate the bias attributable to unmeasured confounding, we hereafter evaluate each method against its own "true" values. In other words, for the DIC method, our target estimand is the disparity explained (or unexplained) after accounting for intermediate confounders $X$; for the KOB decomposition, it is the disparity explained without accounting for baseline covariates $C$; and for the CDA, it is the disparity explained after accounting for baseline covariates while including intermediate confounders in the total disparity. For each scenario, we compare the three methods using DAGs (Figure 3) and simulation results (Tables 2 and 3).

The two scenarios include an unmeasured confounder $U$, which follows a standardized normal distribution ($U \sim N(0, 1)$). The specific data generation procedure is detailed in Supplementary Material B.

### 4.1. Scenario 5

When unmeasured confounding exists in the $X - M$ Relationship, the estimands identified on DAGs by the KOB and CDA frameworks remain unchanged. Figures 3B and 3C show that, despite the presence of unmeasured confounding in the $X - M$ relationship, the pathways for the explained (dotted arrows) and unexplained (bold arrows) portions

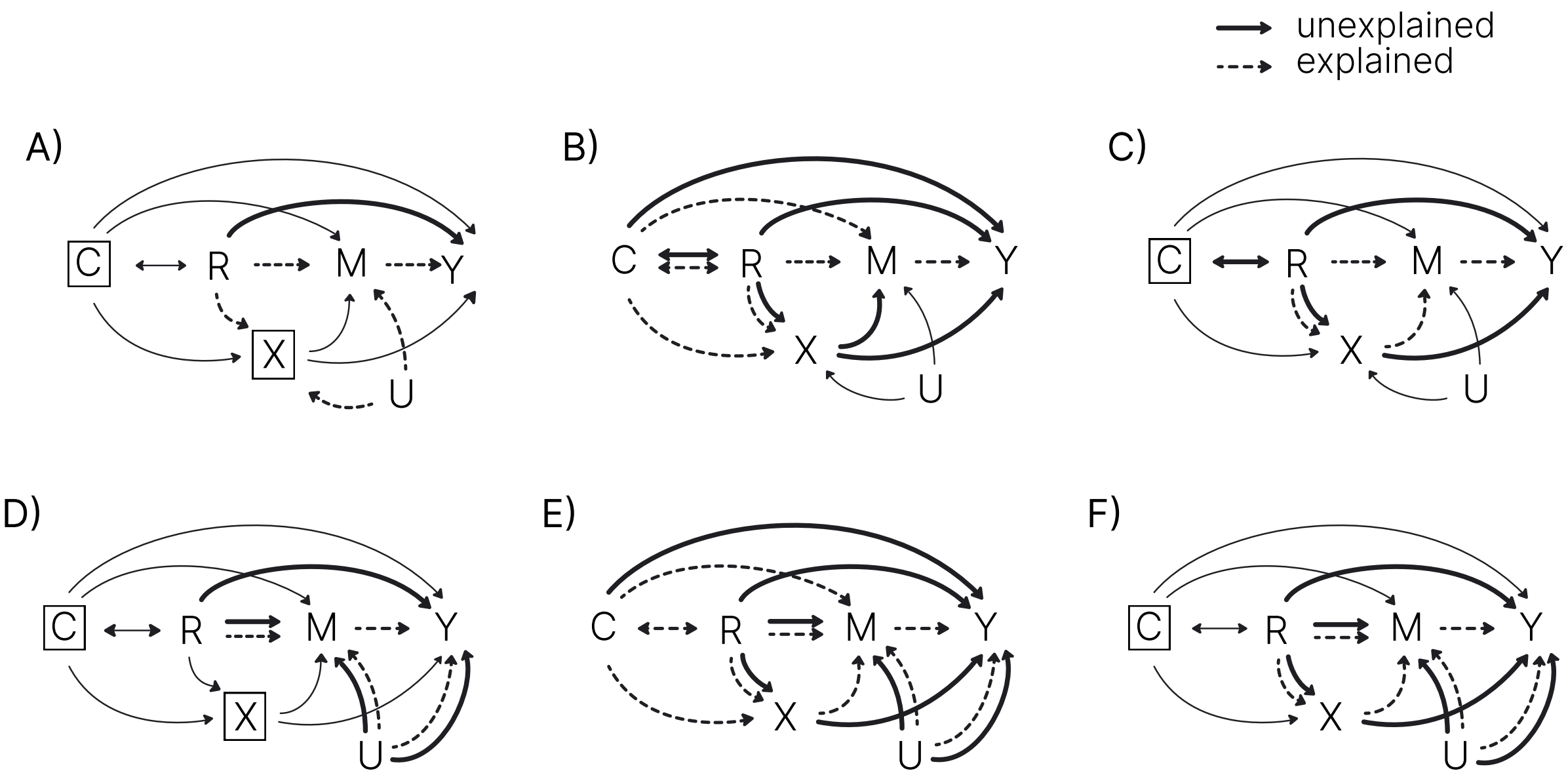

*Figure 3*. DAGs Assuming Omitted Confounders

*Note: 1) A box around a node indicates conditioning on that variable. 2) Dotted arrows indicate the portion of the disparity attributable to differences in the risk factor M. 3) Bold arrows indicate the portion of the disparity not explained by differences in the risk factor M.*

remain the same as those in Figures 1B and 1C, respectively. This conclusion is also corroborated by the estimates with small relative bias (< 3% of the true value) obtained from these methods in Table 2.

In contrast, the total disparity and explained disparity mapped via the DIC framework diverge. Figure 3A shows that the portion explained by education $M$ (dotted arrows) includes a spurious pathway connecting between racial group $R$ and the mediator $M$ through the unmeasured confounder $U$ (e.g., neighborhood environment in childhood). This arises because we condition on intermediate confounders $X$, which is a descendant of both $R$ and $U$ ($R \rightarrow \boxed{X} \leftarrow U$), which is called a *collider* (Pearl, 2009). Conditioning on a variable usually blocks pathways through that variable, but conditioning on a collider induces a spurious association between its parent variables.

This spurious pathway through $U$ changes the absolute magnitude of explained disparity, as shown in our simulations (Table 2). For example, the DIC's estimate of the explained disparity estimate is -0.201 (deviating from the true value of -0.251), reflecting the bias introduced by conditioning on intermediate confounders $X$. Consequently, the total disparity estimate is also biased. This bias due to unmeasured confounding in the $X - M$ relationship can be removed if the mediator were randomly assigned. This further validates our methodological guideline that the DIC is inappropriate for observational research where such randomization is absent.

Table 2

*Disparity Estimates Assuming Omitted Confounders in the X–M Relationship*

| | DIC | | KOB | | CDA | |
|---|---|---|---|---|---|---|
| | Estimate [CI] | True | Estimate [CI] | True | Estimate [CI] | True |
| Total Disparity | −2.200 [−2.320, −2.070] | −2.253 | **-2.520** [−2.660, −2.370] | −2.529 | **-2.380** [−2.550, −2.170] | −2.388 |
| Unexplained (Remaining) | **-2.000** [−2.090, −1.880] | −2.002 | **-2.330** [−2.470, −2.200] | −2.331 | **-2.120** [−2.260, −1.960] | −2.123 |
| Explained (Reduction) | −0.201 [−0.275, −0.126] | −0.251 | **-0.193** [−0.277, −0.116] | −0.198 | **-0.261** [−0.400, −0.143] | −0.265 |

*Note:* DIC = Difference-in-Coefficients; KOB = Kitagawa–Oaxaca–Blinder (Blacks' coefficients as the reference); CDA = Causal Decomposition Analysis. "Estimate [CI]" shows the point estimate and its 95% empirical confidence interval. Estimates with small relative bias (less than 3% away from the true value) are bolded.

### 4.2. Scenario 6

As expected, the unmeasured confounding in the $M - Y$ relationship introduces bias into both the explained and unexplained portions of the disparity across these estimation methods. As shown in Figures 3D-F, both the explained portion (dotted arrows) and the unexplained portion (bold arrows) include pathways through $U$. Consequently, the explained and unexplained portions of disparity estimates obtained from all methods in Table 3 are biased.

However, the total disparity estimates from all methods in Table 3 are closely matched to the true values. This is because the effects of the pathways through $U$ on the explained portion are exactly offset by their effects on the unexplained portion. Therefore, the total disparity estimate is robust to unmeasured mediator-outcome confounding.

The results indicate that unmeasured mediator-outcome $(M - Y)$ confounding cannot be resolved simply by choosing a different method. Rather, it requires a specialized

Table 3

*Disparity Estimates Assuming Omitted Confounders in the M–Y Relationship*

| | **DIC** | | **KOB** | | **CDA** | | **Adj. CDA** | |
|---|---|---|---|---|---|---|---|---|
| | Est. | True | Est. | True | Est. | True | Est. | True |
| **Total Disparity** | **−2.260**<br>[−2.460, −2.080] | −2.248 | **−2.520**<br>[−2.710, −2.350] | −2.524 | **−2.380**<br>[−2.610, −2.160] | −2.383 | **−2.392**<br>[−2.631, −2.146] | −2.383 |
| **Unexplained** | −1.810<br>[−1.940, −1.700] | −1.999 | −2.170<br>[−2.330, −2.020] | −2.328 | −1.930<br>[−2.160, −1.720] | −2.119 | **−2.126**<br>[−2.347, −1.924] | −2.119 |
| **Explained** | −0.445<br>[−0.579, −0.321] | −0.249 | −0.348<br>[−0.486, −0.232] | −0.197 | −0.454<br>[−0.690, −0.254] | −0.264 | **−0.267**<br>[−0.464, −0.064] | −0.264 |

*Note:* DIC = Difference-in-Coefficients; KOB = Kitagawa–Oaxaca–Blinder (Blacks' coefficients as the reference); CDA = Causal Decomposition Analysis; Adj. CDA = Adjusted CDA. Each "Est." cell shows the point–estimate above its empirical 95% CI; Estimates with small relative bias (less than 3% away from the true value) are bolded.

tool to assess robustness. To address this, we utilize the sensitivity analysis proposed in Park, Kang, Lee, and Ma (2023). The procedure identifies the potential bias from an unmeasured confounder ($U$) using two sensitivity parameters based on $R^2$ values: 1) the strength of the association between the unmeasured confounder $U$ and the outcome $Y$ given $R$, $X$, $M$, and $C$, and 2) the strength of the association between the unmeasured confounder $U$ and the mediator $M$ given $R$, $X$, and $C$. If the association between $U$ and both the mediator and the outcome are sufficiently strong, the estimated disparity reduction or the remaining may be nullified or lose statistical significance. Once plausible values for these two sensitivity parameters are specified, the sensitivity analysis recalculates the decomposition to produce bias-adjusted estimates. As shown in Table 3, when the true sensitivity parameters are given, sensitivity analysis produces the bias-adjusted estimates.

Unfortunately, sensitivity parameters cannot be directly estimated because $U$ is unmeasured. Therefore, researchers must rely on their subjective judgment to specify plausible values for these sensitivity parameters. As an alternative, Cinelli and Hazlett (2020) propose a benchmarking strategy that uses the effects of existing confounders to calibrate and interpret the likely strength of unmeasured confounders, which we demonstrate in Section 6.

## 6. Case Study

In this section, we apply three distinct methods to the HRS dataset to quantify the extent to which educational attainment explains racial disparities in later-life cognition. Following the diagnostic guideline in Figure 2, we identify two primary challenges: 1) educational attainment is not randomly assigned and 2) age is correlated with race, with Black participants being, on average, 4.4 years younger than White participants ($p < .001$). In addition, age is significantly associated with both educational attainment ($r = -0.051$) and cognitive function ($r = -0.242$). These conditions necessitate CDA to ensure valid identification. However, we will compare estimates across methods to highlight specific biases that arise when inappropriate methods were used.

**Comparison of Methods.** As shown in Table 4, the total disparity estimates vary substantially across methods, ranging from -1.816 (KOB) to -2.409 (CDA). The KOB approach likely underestimates the Black-White difference because it fails to adjust for age ($C$). It is a critical factor given that older adults from racial minority groups are less likely to participate in cognitive assessments. Conversely, while the DIC approach controls for age, it also controls for early childhood adversities ($X$). Because the Black-White difference is then measured among individuals with same levels of early childhood adversities, the total disparity is likely be attenuated compared to the CDA estimate (DIC: -2.031 vs. CDA: -2.409).

Notably, the explained portion estimated via the DIC method is approximately zero. This result implies that within this model's specification, the Black-White cognitive differential is mainly through early childhood adversities, such as childhood SES. Because these adversities strongly predict their educational attainment, the indirect pathway though education (independent of childhood factors) appears negligible. This highlights the risk of using the DIC method to observational, non-randomized settings.

Lastly, while the explained component estimated via the KOB method is smaller than the CDA estimate (KOB: -0.269 vs. CDA: -0.362), the resulting percentages of the disparity explained are remarkably similar at 14.8% and 15.0%, respectively. This suggests that the bias from age attenuates both the total disparity and the explained portion in a roughly proportional manner, leading to a spurious consistency in the relative percentages ($\frac{-0.269}{-1.816} \times 100 \approx \frac{-0.362}{-2.409} \times 100$). However, this proportional scaling is case-specific. In cases where baseline covariates influence the total disparity and the mediating pathways in an opposite direction, the percentage explained could diverge significantly, leading to highly misleading conclusions.

Table 4

*Comparison of Decomposition Methods for Racial Disparities in Cognition*

| **Method** | **Total Disparity** Estimate [95% CI] | **Explained (Reduction)** Estimate [95% CI] | **Unexplained (Remaining)** Estimate [95% CI] | **% Exp.** |
|---|---|---|---|---|
| DIC | -2.031 [-2.218, -1.846] | -0.000 [-0.053, 0.052] | -2.031 [-2.208, -1.853] | 0.0% |
| KOB | -1.816 [-2.001, -1.616] | -0.269 [-0.343, -0.201] | -1.546 [-2.175, -1.785] | 14.8% |
| CDA | -2.409 [-2.806, -2.098] | -0.362 [-0.672, -0.224] | -2.047 [-2.239, -1.825] | 15.0% |

*Note: DIC = Difference in Coefficients; KOB = Kitagawa-Oaxaca-Blinder; CDA = Causal Decomposition Analysis.*

**Sensitivity Analysis and Covariate Benchmark.** Given the observational nature of the HRS data, the CDA estimates remain vulnerable to potential unmeasured confounding, such as neighborhood environment. We therefore assess the sensitivity of the disparity reduction estimate by varying two sensitivity parameters that represent the strength of the association between unmeasured confounding and both educational attainment and cognition, after controlling for existing covariates.

As illustrated in Figure 4, these parameters are mapped onto the $X$-axis (partial $R^2$ of the unmeasured confounding with the outcome) and the $Y$-axis (partial $R^2$ of the unmeasured confounding with the mediator). The bold contour line identifies the combinations of confounding strength required to reduce the reduction estimate to zero. Correspondingly, the dotted contour line represents the points at which the estimate is no longer significant at the 95% confidence level. This sensitivity analysis was conducted using the `causal.decomp` R package. For researchers interested in alternative frameworks, the `disparity` R package is available at https://github.com/laiweisociology/disparity.

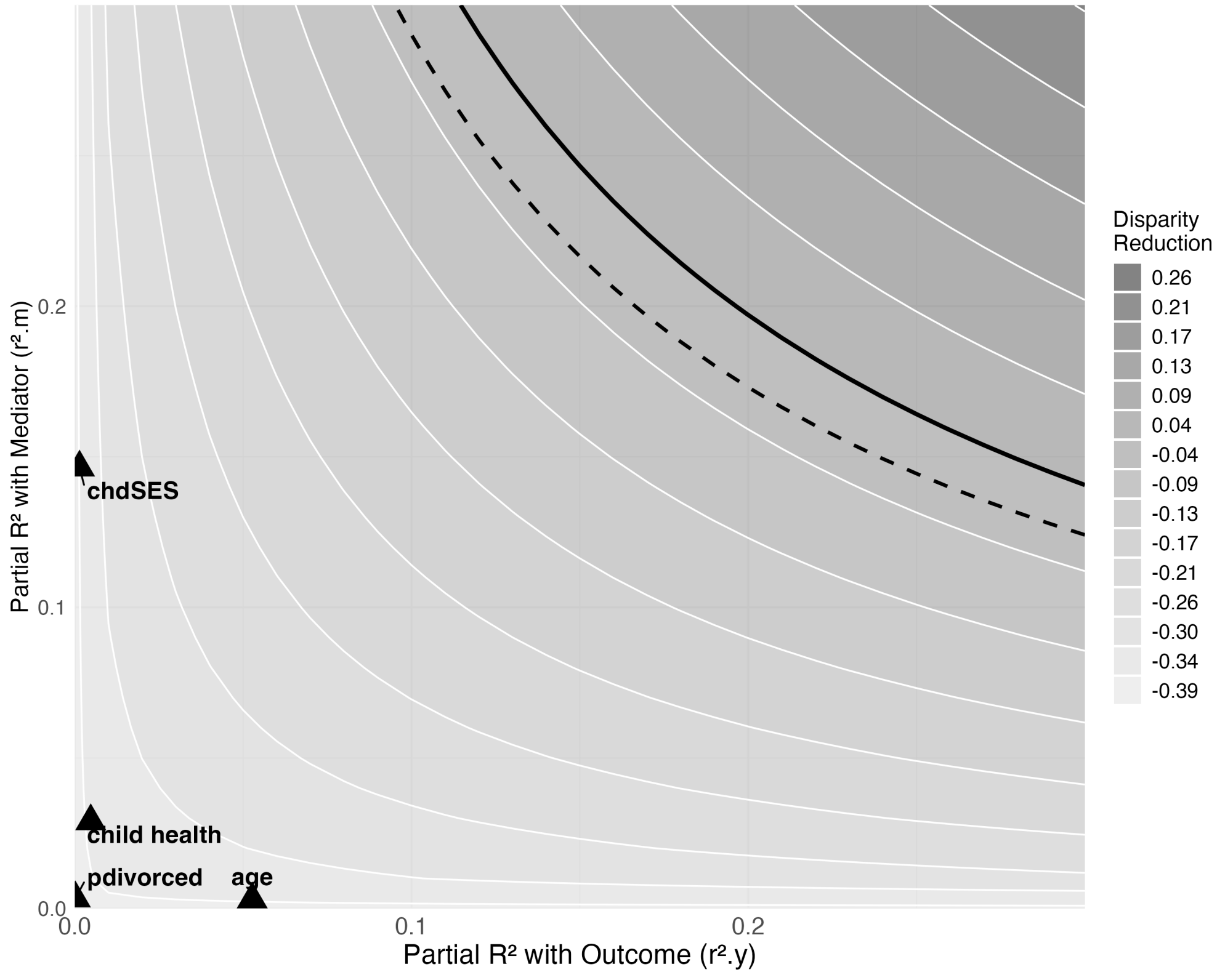

*Figure 4*. Sensitivity of Disparity Reduction Estimates and Covariate Benchmarks

*Note: 1) Bold and dotted contour lines: points where the estimate becomes zero (bold) or is no longer statistically significant (dotted). 2) chdSES= childhood SES, pdivorced = family instability, child health= childhood health.*

To evaluate whether these combinations of confounding strength are plausible, we employ the covariate benchmarking method by Cinelli and Hazlett (2020). This approach calibrates the potential confounding strength by using existing covariates as empirical references. In Figure 4, these benchmarks are represented as triangles. Notably, all observed covariates are well below the dotted contour line, indicating that unmeasured confounding to nullify our results would need to be substantially more influential than the current predictors. For example, even if an unmeasured confounder, such as neighborhood environment in childhood, has an association with the outcome as strong as that of age ($R^2 = 0.053$), and an association with the mediator as strong as that of childhood SES ($R^2 = 0.146$), the disparity reduction estimate would remain negative and statistically significant. This benchmarking method demonstrates that our disparity reduction estimate is robust to omitted variable bias.

## 7. Implications for Designing and Analyzing Preventive Interventions

This study compares widely-used decomposition methods and offers best-practice recommendations for applied researchers. Currently, method selection is often driven by familiarity or software accessibility rather than causal logic. This study emphasizes the importance of selecting a method based on the specific research setting and confounding structure, demonstrating that the use of an inappropriate method can lead to significant estimation bias.

In settings where the mediator is randomly assigned, the standard DIC method is generally sufficient. For example, in a school-based preventive program designed to mitigate baseline racial disparities in adolescent substance use, the standard DIC can effectively estimate both the baseline total disparity and the remaining post-intervention disparity. The baseline racial disparities in the outcome can be obtained directly from the pre-intervention data, while the remaining post-intervention disparity can be obtained by fitting a treatment-adjusted regression model. Note, however, that directly comparing the baseline total disparity to the remaining post-intervention disparity requires the pre- and post-intervention outcomes to be placed in the same scale through test equating.

Conversely, in observational settings where the mediator is not randomly assigned, the mediator-outcome relationship is highly susceptible to confounding by both baseline covariates and intermediate confounders. In these contexts, we recommend using the modified KOB or CDA.

The choice between these methods depends on the researcher's objectives. The modified KOB is designed for descriptive purposes (Opacic et al., 2025). For example, it is ideal when mediators are variables that cannot be directly intervened on, such as parental education. In these cases, the research question focuses on how much of an observed gap is explained by parental education. Yet, even though the objective is descriptive, the assumption of no omitted confounding is still required, making sensitivity analysis essential.

In contrast, CDA is designed for prescriptive purposes, addressing “what-if” scenarios such as, what if racial groups were equally exposed to the preventive program?). Consequently, the mediators evaluated under CDA must be policy-relevant, malleable factors that can actually be targeted and changed. Ultimately, CDA can leverage observational data to identify critical boundary conditions, for instance, clarifying whether school-based or family-focused interventions would be more effective in terms of reducing racial disparities.

However, CDA estimates should not be interpreted as definitive evidence of an intervention’s actual efficacy. Rather, researchers and practitioners should view them as an informative, relatively robust evidence base for implementing or designing efficacy trials or preventive strategies. That is because, while the CDA framework can incorporate a sensitivity analysis to mitigate potential violations of the no omitted confounding assumption, other assumptions, such as consistency or positivity, remain vulnerable. If these assumptions are not met, the estimated causal effects of hypothetical interventions may still be compromised.

Our review is limited to the context of decomposing social disparities. If researchers seek to understand the mechanisms underlying the effect of a randomized treatment, such as the job training program as in Imai et al. (2010) where the mediator is a psychological state like ‘job search self-efficacy’ that is not easily intervened upon directly, natural indirect effects are more appropriate. In this context, researchers can conceptualize counterfactual, such as an individual’s self-efficacy level under treatment vs. control conditions. Furthermore, if self-efficacy is measured immediately following the intervention, the ‘no intermediate confounding’ assumption becomes more plausible. In this specific context, the DIC approach also provides an unbiased estimate. Future research should further compare natural indirect effects with traditional mediation frameworks within these experimental contexts.